# Dynamic Energy Flow Analysis of Integrated Electricity and Gas Systems: A Semi-Analytical Approach

Zhikai Huang, *Student Member, IEEE,* Shuai Lu, Wei Gu, *Senior Member,* Ruizhi Yu, Suhan Zhang, Yijun Xu, *Senior Member, IEEE,* Yuan Li

*Abstract*—Ensuring the safe and reliable operation of integrated electricity and gas systems (IEGS) requires dynamic energy flow (DEF) simulation tools that achieve high accuracy and computational efficiency. However, the inherent strong nonlinearity of gas dynamics and its bidirectional coupling with power grids impose significant challenges on conventional numerical algorithms, particularly in computational efficiency and accuracy. Considering this, we propose a novel non-iterative semi-analytical algorithm based on differential transformation (DT) for DEF simulation of IEGS. First, we introduce a semi-discrete difference method to convert the partial differential algebraic equations of the DEF model into ordinary differential algebraic equations to resort to the DT. Particularly, by employing spatial central difference and numerical boundary extrapolation, we effectively avoid the singularity issue of the DT coefficient matrix. Second, we propose a DT-based semi-analytical solution method, which can yield the solution of the DEF model by recursion. Finally, simulation results demonstrate the superiority of the proposed method.

## I. INTRODUCTION

### A. Background

INTEGRATED energy systems have become crucial for addressing growing energy demands and sustainability challenges through multi-energy coordination [1]. The synergy between natural gas networks (NGN) and electric power systems (EPS) significantly enhances energy efficiency and reduces emissions [2], yet their complex interdependencies pose operational risks that demand rigorous analysis. The emergence of power-to-gas (P2G) technology is transforming NGN-EPS interactions from unidirectional to bidirectional exchanges [3], further adding complexity to integrated electricity and gas system (IEGS) operations.

The dynamic energy flow (DEF) analysis is a critical foundation for evaluating the operational security of IEGS, requiring large-scale nonlinear partial differential algebraic equations (PDAEs). While numerical approaches like finite difference methods (FDMs) [3, 4] and characteristic line methods [5] have been explored for the DEF analysis, they face computational bottlenecks, including intensive computational cost and convergence issues because of the nonlinear NGN and EPS equations. In light of this, we develop a semi-analytical (SA) algorithm that features high accuracy and low computational cost for the DEF analysis of IEGS, offering an efficient tool for the operational analysis of IEGS.

### B. Literature Review

The energy flow analysis of IEGS can be divided into the steady one and the dynamic one based on the energy flow transport models it uses. The steady energy flow analysis uses algebraic equations (AEs) to describe the gas flows in the pipelines by ignoring the dynamic evolution process, focusing on the long-term (usually above hours) operational conditions of the systems [6]. Comparatively, the DEF analysis uses partial differential equations (PDEs) to describe the gas flow dynamics along the pipelines, focusing on the short-term (from seconds to minutes) operational conditions of the systems [7].

This work focuses on the DEF analysis problem of IEGS, which is typically large-scale nonlinear PDAEs because of the occurrence of the nonlinear gas dynamics in NGN and the nonlinear power flow in EPS. To solve the DEF model, the traditional numerical methods consist of two steps: 1) Discretizing the PDEs into AEs using methods such as FDMs and finite volume methods; 2) Solving the nonlinear AEs using iterative methods, typically including Newton's method, Quasi-Newton method, and Levenberg-Marquardt Method. For example, in [8], the implicit Euler difference schemes are used to discretize the PDEs of gas flow, and in [4], the centered-difference form in space and the fully implicit algorithm in time are proposed for the gas flow equations and then compared with other difference schemes under Newton's method. Although the numerical methods are easy to use, they have some inherent drawbacks. First, in the discretization step, to ensure stability and accuracy, the temporal and spatial steps cannot be too large and need to satisfy some specific constraints, which usually introduce many variables and equations, greatly increasing the problem scale. Second, in solving the nonlinear AEs, it is very challenging to select a starting point that is guaranteed to be in the convergence region [9], which is still an open problem in the field of nonlinear system computation.

Considering the shortcomings of the numerical methods, the SA approaches have received some attention in recent years, such as the differential transformation (DT) and the holomorphic embedded (HE) methods [10-12], featuring the rigorousness of analytical methods and the flexibility of numerical methods. Particularly, the SA approaches do not require iterative calculation to solve nonlinear problems and thus outperform in calculation accuracy and convergence. In the field of energy systems, the SA approach was first utilized to solve the power flow model with AEs and electromechanical transient simulation with ordinary differential algebraic equations (ODAEs) [13, 14]. By contrast, designing an efficient SA-based algorithm for the DEF model of IEGS is more challenging since it is a nonlinear PDAE model.

Recently, the SA approaches have also been explored in the field of integrated energy systems. Yu *et al.* [15] first investigate the DT-based SA approach for solving the DEF model of integrated heat and electricity systems. Zhang *et al.* [16] and Huang *et al.* [17] explore the HE-based SA approach for solving the DEF model of IEGS, in which the treatment of difference schemes and boundary conditions may cause some problems. Specifically, their approach encounters singularities in the coefficient matrix of the equations when addressing loop networks or multiple gas source problems in NGN due to inconsistent pressures at pipeline tail converging on the same node. In [16], an assumption of equal pressures at the head and tail of the last segment of these pipelines is made, which may introduce computational errors. Overall, the SA approach for the DEF models is still in the initial stage. The research gaps include: 1) How to efficiently deal with the PDAE model of NGN in the SA approach is still unsolved; 2) it remains unclear how to exploit the special structure of the DEF model in IEGS to improve the computing performance.

*C. Contributions*

To bridge the aforementioned research gaps, we propose a novel SA algorithm for the DEF model of IEGS, which features higher computation efficiency and robustness while maintaining high accuracy. Case studies verify the effectiveness of the proposed method. The main contributions are summarized as follows.

(1) We propose a novel non-iterative DT-based algorithm for the DEF model of IEGS. Compared with the numerical method, it avoids the iterative calculation of the equation and greatly reduces the number of matrix inversions.

(2) We propose a spatial difference scheme and a numerical boundary construction method to convert the PDE of gas dynamics into ODAEs to resort to the SA approach, solving the coefficient matrix singularity problem in the SA approach and featuring the 2nd-order spatial precision.

(3) We propose an adaptive window control technique for the DT-based DEF calculation, which uses estimated truncation errors to adjust the size of each time window for higher robustness.

## II. PROBLEM DESCRIPTION

In this section, we first present the DEF model of IEGS. Second, the initial and boundary conditions for solving the DEF model are formulated. Finally, the classical FDM-based numerical algorithm for the DEF model is briefly introduced.

*A. Dynamic Energy Flow Model*

The DEF model of IEGS includes the NGN model, the EPS model, and the coupling components model, which are introduced separately in the following.

*1) Natural gas network*

The NGN transports natural gas from source nodes to load nodes through pipelines. The NGN model describes the gas dynamics in the pipelines and the mass conservation and pressure consistency at the nodes. The schematic of the NGN model is given in Fig. 1.

The gas dynamics in the pipeline can be described by the mass conservation and momentum conservation equations [18]. In this work, we adopt the following assumptions widely used in existing work [3, 19] to model the gas dynamics including: 1) The pipeline is horizontal; 2) The gas transmission in pipelines is a constant temperature process; 3) The convection in the pipeline is negligible. Under these assumptions, the gas dynamics in the pipeline can be described by the continuity equation (1a) and momentum equation (1b), as:

$$\frac{\partial \pi_j^{pl}}{\partial t} + \frac{c^2}{S_j}\frac{\partial m_j^{pl}}{\partial l} = 0, \quad (1a)$$

$$\frac{\partial m_j^{pl}}{\partial t} + S_j \frac{\partial \pi_j^{pl}}{\partial l} + \frac{\lambda_j c^2 m_j^{pl}|m_j^{pl}|}{2 D_j S_j \pi_j^{pl}} = 0, \quad (1b)$$

wherein $j = 1,2,\cdots,J$ is the index of the pipeline; $J$ is the number of pipelines in NGN; $S_j$ is the cross-sectional area of pipeline $j$, m$^2$; $\lambda_j$ is the friction factor of pipeline $j$; $D_j$ is the diameter of pipeline $j$, m; $c$ is the sound velocity in gas, m/s; $\pi_j^{pl}$ is the gas pressure of pipeline $j$, Pa; $m_j^{pl}$ is mass flow of pipeline $j$, kg/s.

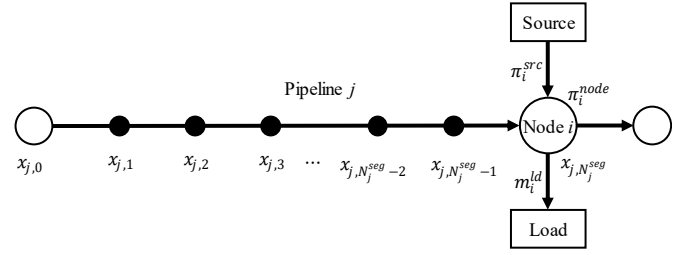

Fig. 1. Diagram of the model of NGN.

Here, since $\pi_j^{pl}$ and $m_j^{pl}$ are spatially distributed along the pipeline $j$, as indicated in Fig. 1, in the following, we use $m_j^{pl}(l,t)$ and $\pi_j^{pl}(l,t)$, $l \in [0,L_j], t \in [0,T]$ to denote their values in the location $l$ and time $t$ to facilitate the modeling of nodes, wherein $L_j$ is the length of pipeline $j$ and $T$ is the simulation duration, respectively. Especially we use $m_j^{pl}(0,t)$ and $\pi_j^{pl}(0,t)$ (or $m_j^{pl}(L_j,t)$ and $\pi_j^{pl}(L_j,t)$) to denote the values at the head (or tail) of pipeline $j$ at time $t$.

Next, we introduce three matrixes to describe the mass conservation and pressure consistency at the nodes, including $K^{in} \in \mathbb{R}^{I \times J}$, $K^{out} \in \mathbb{R}^{I \times J}$, and $K^{cmp} \in \mathbb{R}^{I}$, wherein $I$ is the number of nodes in NGN. Their elements are defined as:

$$K_{ij}^{in} = \begin{cases} 1, & \text{If the gas in pipeline } j \text{ flows into node } i \\ 0, & \text{Otherwise} \end{cases},$$

$$K_{ij}^{out} = \begin{cases} 1, & \text{If the gas in pipeline } j \text{ flows out of node } i \\ 0, & \text{Otherwise} \end{cases},$$

$$K^{cmp} = \begin{cases} k_i^{cmp}, & \text{If node } i \text{ has the compressor} \\ 1, & \text{Otherwise} \end{cases},$$

wherein $K^{in}$ and $K^{out}$ describe the connection relationship between pipelines and nodes, respectively; $K^{cmp}$ denotes the existence of compressors at nodes; $k_i^{cmp}$ is the pressure ratio of the compressor at node $i$.

Then, the nodal mass conservation equations are as:

$$m^{nd}(t) = K^{out} m^{pl}(0,t) - K^{in} m^{pl}(L,t), \quad (1c)$$

wherein $m^{nd}(t) \in \mathbb{R}^{I}$ is the vectors of injection mass flow of nodes at time $t$; $m^{pl}(0,t) = [m_1^{pl}(0,t) \ m_2^{pl}(0,t) \ \cdots$ $m_J^{pl}(0,t)]^\top \in \mathbb{R}^J$ and $m^{pl}(L,t) = [m_1^{pl}(L_1,t)\ m_2^{pl}(L_2,t) \cdots m_J^{pl}(L_J,t)]^\top \in \mathbb{R}^J$ are the vectors of the gas mass flow at the head and tail of all the pipelines, respectively; $L = [L_1\ L_2 \cdots L_J] \in \mathbb{R}^J$.

The pressure consistency equations are as:

$$(K^{in})^\top \pi^{nd}(t) = \pi^{pl}(L,t), \quad (1d)$$

$$(K^{out})^\top \mathrm{diag}(K^{cmp}) \pi^{nd}(t) = \pi^{pl}(0,t), \quad (1e)$$

wherein $\pi^{nd}(t) \in \mathbb{R}^I$ is the vectors of gas pressure of nodes at time $t$; $\pi^{pl}(0) = [\pi_1^{pl}(0,t)\ \pi_2^{pl}(0,t) \cdots \pi_J^{pl}(0,t)]^\top \in \mathbb{R}^J$; $\pi^{pl}(L,t) = [\pi_1^{pl}(L_1,t)\ \pi_2^{pl}(L_2,t) \cdots \pi_J^{pl}(L_J,t)]^\top \in \mathbb{R}^J$ are the vectors of the gas pressure flow at the head and tail of all the pipelines, respectively.

*2) Electrical power system*

In this work, we use the steady-state power flow model in a rectangular coordinate system for the EPS model, as follows:

$$p = \mathrm{diag}(e)(Ge - Bf) + \mathrm{diag}(f)(Be + Gf), \quad (1f)$$

$$q = \mathrm{diag}(f)(Ge - Bf) - \mathrm{diag}(e)(Be + Gf), \quad (1g)$$

$$\mathrm{diag}(e)e + \mathrm{diag}(f)f = \mathrm{diag}(U)U, \quad (1h)$$

wherein $G$ and $B$ are the conductance and susceptance matrices; $p \in \mathbb{R}^{N^{PQ}+N^{PV}}$ and $q \in \mathbb{R}^{N^{PQ}}$ are the vectors of active and reactive power injected into the buses, respectively; $U \in \mathbb{R}^{N^{PV}}$ is the vector of bus voltage magnitude on PV buses; $e \in \mathbb{R}^{N^{bus}}$ and $f \in \mathbb{R}^{N^{bus}}$ are the vectors of real and imaginary parts of the voltage on buses, respectively; $N^{PQ}$, $N^{PV}$, and $N^{bus}$ are the numbers of the PQ, PV, and total buses, respectively.

*3) Coupling Components*

Typically, the coupling components between NGN and EPS include electricity-driven compressors, gas-fired turbines (GT), and P2G equipment. For the electricity-driven compressor with known pressure ratios at node $i$ in NGN and bus $b$ in EPS, we use the linear model to calculate the power consumption [20], as:

$$\begin{aligned} K_{b,i}^C p_b &= -K_i^{out} m^{pl}(0,t) \quad \forall (i,b) \in S^C \\ q_b &= \tan\theta_b p_b \end{aligned}, \quad (1i)$$

wherein $K_{b,i}^C$ is the efficiency of the gas compressor; $\theta_b$ is the phase difference angle of the compressor; $p_b$ and $q_b$ are the active and reactive power injection for the bus where the compressor is located, respectively; vector $K_i^{out}$ is the $i$th row of the matrix $K^{out}$; $S^C$ is the set of NGN nodes and EPS bus labels where all compressors are located.

For the GT that is located at node $i$ in NGN and supplies power at bus $b$ in EPS, the generated power is calculated as [3]:

$$p_b = -K_{b,i}^{GT} m_i^{nd} \quad \forall (i,b) \in S^{GFU}, \quad (1j)$$

wherein $K_{b,i}^{GT}$ is the conversion efficiency of gas into electricity; $p_b$ is the active power injection for the bus where the GT is located; $m_i^{nd}$ is the gas injection for the node where the GT is located; $S^{GT}$ is the set of NGN nodes and EPS bus labels where all GTs are located.

For the P2G that is located as bus $b$ in EPS and supplies gas at node $i$ in NGN, the gas output can be calculated as [16]:

$$\begin{aligned} m_i^{nd} &= -K_{b,i}^{P2G} p_b \quad \forall (i,b) \in S^{P2G} \\ q_b &= \tan\theta_b p_b \end{aligned}, \quad (1k)$$

wherein $K_{b,i}^{P2G}$ is the conversion efficiency of electricity into gas; $\theta_b$ is the phase difference angle of the P2G; $p_b$ and $q_b$ are the active and reactive power injection for the bus where the compressor is located, respectively; $m_i^{nd}$ is the gas injection for the node where the P2G is located; $S^{P2G}$ is the set of NGN nodes and EPS bus where the P2Gs are located.

*B. Initial and Boundary Conditions*

For the DEF model, the initial and boundary conditions are the prerequisites for the uniqueness of the solution. The initial conditions provide the initial values of states for solving the (partial) differential equations, while the boundary conditions offer state constraints at the boundaries of the systems. In the context of IEGS, the initial conditions define the initial states of the gas pressure and mass flow of the gas pipelines at the beginning of the simulation, as:

$$\pi_j^{pl}(l,0) = \pi_j^{init}(l) \quad \forall l \in [0,L_j], j=1,2,\cdots,J, \quad (1l)$$

$$m_j^{pl}(l,0) = m_j^{init}(l) \quad \forall l \in [0,L_j], j=1,2,\cdots,J, \quad (1m)$$

wherein $\pi_j^{init}$ and $m_j^{init}$ are the initial values of pipeline pressure and mass flow, respectively.

The boundary conditions include the ones of NGN, EPS, and coupling components. For NGN, the boundary conditions include the gas pressure of gas source nodes and the mass flow of load nodes, as:

$$\pi_i^{nd}(t) = \pi^{src}(t) \quad \forall i \in S^{src}, \quad (1n)$$

$$m_i^{nd}(t) = m^{ld}(t) \quad \forall i \in S^{ld}, \quad (1o)$$

wherein $\pi^{src} \in \mathbb{R}^{N^{src}}$ is the known vector of pressure of source nodes; $m^{load} \in \mathbb{R}^{N^{ld}}$ is the known vector of the mass flow of load nodes; $N^{src}$ and $N^{ld}$ are the number of source and load nodes in NGN, respectively; $S^{src}$ and $S^{ld}$ are the set of source and load nodes in NGN, respectively.

For EPS, the boundary conditions include the injected active power $p^{PV}(t)$ and voltage amplitude $U^{PV}(t)$ of PV buses, the injected active power $p^{PV}(t)$ and reactive power $q^{PV}(t)$ of PQ buses, and the real parts $e^{slk}(t)$ and imaginary $f^{slk}(t)$ of the voltage of the slack bus, as:

$$p_b(t) = p^{PV}(t), U_b(t) = U^{PV}(t) \quad \forall b \in S^{PV}, \quad (1p)$$

$$p_b(t) = p^{PQ}(t), q_b(t) = q^{PQ}(t) \quad \forall b \in S^{PQ}, \quad (1q)$$

$$e_b(t) = e^{slk}(t), f_b(t) = f^{slk}(t) \quad \forall b \in S^{slk}, \quad (1r)$$

wherein $S^{PV}$, $S^{PQ}$, and $S^{slk}$ are the sets of labels for PV, PQ, and slack bus, respectively.

Furthermore, the boundary conditions for coupling components hinge on their respective control strategies. Electricity-driven compressors, serving as PQ buses in EPS, have their active and reactive power determined by the mass flow through them in NGN via the equation (1i), which remains unknown prior to calculation. GTs serving as the slack bus in EPS and the load node in NGN have known real and imaginary voltage components. However, their unknown power must be derived through power flow calculations, while the mass flow of consumed gas is solved using the equation (1j) based on their active power. GTs serving as PV buses in EPS and load nodes in NGN have known active power and voltage amplitude. The mass flow of consumed gas is calculated using the equation (1j) based on their active power, and other unknowns are resolved through power flow calculations. P2Gs serving as PQ buses in EPS and negative load nodes in NGN have known active and reactive power. The mass flow of output gas is determined by the equation (1k) based on their active power, with other unknowns resolved via power flow calculations.

### C. FDM-Based Numerical Algorithm

The FDM-based numerical algorithm for the DEF model includes two steps: choosing a difference scheme to discretize the PDEs into AEs and using the iterative method to solve the system of AEs. In existing work, the commonly-used FDMs for the PDEs of gas flow include the implicit Euler [8] and implicit central method [21], separately as follows:

$$\frac{\partial x_j}{\partial l} \approx \frac{x_{j,n+1,\tau+1} - x_{j,n,\tau+1}}{\Delta l}$$
$$\frac{\partial x_j}{\partial t} \approx \frac{x_{j,n+1,\tau+1} - x_{j,n+1,\tau}}{\Delta t}, \quad (2a)$$
$$x_j \approx x_{j,n+1,\tau+1}$$

$$\frac{\partial x_j}{\partial l} \approx \frac{x_{j,n+1,\tau+1} - x_{j,n,\tau+1} + x_{j,n+1,\tau} - x_{j,n,\tau}}{2\Delta l}$$
$$\frac{\partial x_j}{\partial t} \approx \frac{x_{j,n+1,\tau+1} - x_{j,n+1,\tau} + x_{j,n,\tau+1} - x_{j,n,\tau}}{2\Delta t} \quad . \quad (2b)$$
$$x_j \approx \frac{1}{4}\left(x_{j,n+1,\tau+1} + x_{j,n,\tau+1} + x_{j,n+1,\tau} + x_{j,n,\tau}\right)$$

wherein $\Delta t$ and $\Delta l$ indicate the temporal and spatial steps, respectively; $\tau = 0,1,\cdots,N^{time}$ and $n = 0,1,\cdots,N_j^{seg}$ are the indexes of the temporal and spatial segments, respectively; $N^{time}$ is the number of moments to be simulated; $N_j^{seg}$ is the number of differential segments of pipeline $j$.

The implicit Euler method has 1st order precision with the truncation error $O(\Delta t + \Delta x)$. The implicit central method has 2nd order precision with the truncation error $O(\Delta t^2 + \Delta x^2)$, while it has large oscillations.

Using the FDM for (1a) and (1b), we can convert the PDEs into AEs. Then, by combining them with the equations (1c)-(1r), we can get the AEs system for the DEF analysis. Next, we need to solve the nonlinear AEs, for which the classical method is Newton's method.

## III. DIFFERENTIAL TRANSFORMATION OF DEF MODEL

In this section, we first introduce the principles and rules of DT. Second, we propose the semi-discrete method to transform the PDEs of gas dynamics into ordinary differential equations (ODEs), based on which the PDAE system of the DEF model is converted into nonlinear ODAEs. Finally, we propose the DT-based SA approach for the DEF model.

### A. Introduction of Differential Transformation

The DT method converts the nonlinear ODAEs to linear AEs. Then, the semi-analytical solutions (SASs) are obtained by calculating the DT coefficients of the unknown variables in the nonlinear ODAEs [11]. We start from the generic ODAEs, as:

$$M\dot{x} = f(x), \quad (3a)$$

wherein $x \in \mathbb{R}^N$ are the variables; $f(x): \mathbb{R}^N \to \mathbb{R}^N$ are (nonlinear) AEs; $M$ is the mass matrix.

In this paper, for the variable $(\cdot)$, we use $\widehat{(\cdot)}[k]$ to denote its $k$ th order DT coefficient. For the equations (3a), the DT coefficients $\hat{x}[k]$ of the variables $x$ are defined as:

$$\hat{x}[k] = \frac{1}{k!}\left[\frac{d^k x(t)}{dt^k}\right]_{t=t_0} \quad k = 0,1,\cdots,K, \quad (3b)$$

wherein $K$ is the selected order of the DT method.

Once we obtain the values of $\hat{x}[k], k = 0,1,\cdots,K$, the approximated SAS of $x$ can be represented as:

$$x(t) \approx \sum_{k=0}^{K} \hat{x}[k] \cdot (t-t_0)^k. \quad (3c)$$

Now, we list the transformation laws used in this paper as follows [10, 11, 13], and interested readers can refer to [11] for a detailed introduction to DT theory: 1) $c \to c\delta[k] \triangleq \begin{cases} c, k = 0 \\ 0, k \geq 1 \end{cases}$,
2) $x(0) \to \hat{x}[0]$, 3) $cx(t) \to c\hat{x}[k]$, 4) $x(t) \pm y(t) \to \hat{x}[k] \pm \hat{y}[k]$, 5) $x(t) \cdot y(t) \to \sum_{m=0}^{k} \hat{x}[m] \cdot \hat{y}[k-m] \triangleq \hat{x}[k] \otimes \hat{y}[k]$,

6) $\frac{1}{x(t)} \to \frac{1}{\hat{x}}[k] \triangleq \begin{cases} \dfrac{1}{x(0)}, k = 0 \\ -\dfrac{\sum_{m=0}^{k-1}\left[\left(\dfrac{1}{\hat{x}}[m] \cdot \hat{x}[k-m]\right)\right]}{\hat{x}[0]}, k \geq 1 \end{cases}$,

7) $\dfrac{y(t)}{x(t)} \to \hat{y}[k] \otimes \dfrac{1}{\hat{x}}[k]$, and 8) $\dfrac{dx}{dt} \to (k+1)\hat{x}[k+1]$.

### B. Semi-Discrete Method of PDE

Obviously, the DT method cannot directly deal with the PDEs of gas dynamics. Hence, we propose to use the semi-discrete method to transform the PDEs into ODEs.

#### 1) Semi-discrete differences

We employ the semi-discrete method [22, 23], also called the method of lines, to transform PDEs into ODEs. Specifically, for the nonlinear PDEs presented in the equations (1a) and (1b), we can reformulate them into a generalized form as follows:

$$\frac{\partial u_j^{pl}}{\partial t} + V_j \frac{\partial u_j^{pl}}{\partial l} = Z_j\left(u_j^{pl}\right) \quad j = 1,2,\cdots,J, \quad (4a)$$

wherein $u_j^{pl} = \begin{bmatrix} \pi_j^{pl} & m_j^{pl} \end{bmatrix}^T$; $Z_j(u_j) = \begin{bmatrix} 0 & -\lambda_j c^2 m_j^{pl}|m_j^{pl}|/\left(2D_j S_j \pi_j^{pl}\right) \end{bmatrix}^T$, and

$$V_j = \begin{bmatrix} 0 & c^2/S_j \\ S_j & 0 \end{bmatrix}.$$

Then, we keep the time derivative and only discretize the space derivative term using a central difference scheme [24], as:

$$\frac{\partial u_j^{pl}}{\partial l} = \frac{u_{j,n+1}^{pl} - u_{j,n-1}^{pl}}{\Delta l} \quad n = 1,2,\cdots,N_j^{seg} - 1. \quad (4b)$$

By substituting (4b) into (4a), we have

$$\frac{du_{j,n}^{pl}}{dt} = V_j \frac{u_{j,n+1}^{pl} - u_{j,n-1}^{pl}}{\Delta l} + Z_j\left(u_{j,n}^{pl}\right) \quad n = 1,2,\cdots,N_j^{seg} - 1. \quad (4c)$$

Also, the initial conditions (1l) and (1m) become

$$u_{j,n}^{pl}(0) = u_j^{pl}(n\Delta l, 0) = u_j^{init}(n\Delta l) \quad n = 0,1,\cdots,N_j^{seg}. \quad (4d)$$

**Remark 1.** For pipeline $j$ comprising $N_j^{seg}$ segments, the equations (4b) encompass a total of $2N_j^{seg} - 2$ equations. However, there are $2N_j^{seg} + 2$ variables in these equations. Even after accounting for the two boundary conditions at the head and tail of the pipeline, the system remains underdetermined, with fewer equations than unknowns. This discrepancy arises because the adopted spatial discretization scheme utilizes points on both sides to approximate the spatial partial derivative of the intermediate point state. To address this issue, we propose the numerical boundary conditions.

*2) Numerical boundary conditions*

First, we perform the eigendecomposition of $V_j$ in (4a) as $V_j = v_j \Lambda_j v_j^{-1}$, wherein $v_j = [c/S_j \ -c/S_j; 1 \ 1]$. Then, we have

$$\Lambda_j = v_j^{-1} V_j v_j = \begin{bmatrix} c & 0 \\ 0 & -c \end{bmatrix},$$

Define $w_j^{pl} \triangleq [w_j^{pl,1} \ w_j^{pl,2}]^\top \triangleq v_j^{-1} u_j^{pl} = 1/(2c) [S_j \pi_j^{pl} + cm_j^{pl} \ -S_j \pi_j^{pl} + cm_j^{pl}]^\top$. Then, from (4a) we have:

$$\frac{\partial w_j^{pl}}{\partial t} + \Lambda_j \frac{\partial w_j^{pl}}{\partial l} = v_j^{-1} Z_j (v_j w_j^{pl}). \quad (4e)$$

According to the total differential formula, we have

$$\frac{dw_j^{pl}}{dt} = \frac{\partial w_j^{pl}}{\partial t} + \frac{\partial w_j^{pl}}{\partial l} \frac{dl}{dt}. \quad (4f)$$

If we set the slope of the characteristic line as

$$\frac{dl}{dt} = c,$$

then the first term of the PDE in (4e) can be converted to ODE

$$\frac{dw_j^{pl,1}}{dt} = -\frac{\lambda_j c^2}{4D_j S_j} \frac{m_j^{pl} |m_j^{pl}|}{\pi_j^{pl}}. \quad (4g)$$

We set the spatial step $\Delta x = L_j/N_j^{seg}$ and the temporal step $\Delta t = \Delta l/c$ here, respectively. By Integrating both sides of (4g), along the characteristic line, we have

$$w_j^{pl,1}(L_j, t + n\Delta t) - w_j^{pl,1}(L_j - n\Delta l, t)$$
$$= -\frac{\lambda_j c^2}{4D_j S_j} \int_t^{t+n\Delta t} \frac{m_j^{pl} |m_j^{pl}|}{\pi_j^{pl}} dt \quad n = 0, 1, 2, \cdots. \quad (4h)$$

Given that $w_j^{pl,1}$ is continuous, according to the mean value theorems for integrals, the integral on the right-hand side in (4h) can be expressed as

$$\int_t^{t+n\Delta t} \frac{m_j^{pl} |m_j^{pl}|}{\pi_j^{pl}} dt = h(l_n^*, t_n^*) \cdot n\Delta t, \quad (4i)$$

wherein $L_j - n\Delta l \leq l_n^* \leq L_j$; $t \leq t_n^* \leq t + n\Delta t$; $h(l_n^*, t_n^*) \triangleq m_j^{pl}(l_n^*, t_n^*) |m_j^{pl}(l_n^*, t_n^*)|/\pi_j^{pl}(l_n^*, t_n^*)$.

Hence, (4h) can be written as:

$$w_j^{pl,1}(L_j, t + n\Delta t) - w_j^{pl,1}(L_j - n\Delta l, t) = -\frac{\lambda_j c^2}{4D_j S_j} h(l_n^*, t_n^*) \cdot n\Delta t. \quad (4j)$$

Second, by using the linear interpolation is considered at $(L_j, t + 2L/(cN_i^{seg}))$, we have:

$$w_j^{pl,1}(L_j, t + 2\Delta t) = 2w_j^{pl,1}(L_j, t + \Delta t) - w_j^{pl,1}(L_j, t) + O((\Delta t)^2). \quad (4k)$$

Based on (4j), we get

$$w_j^{pl,1}(L_j, t + 2\Delta t) - w_j^{pl,1}(L_j - 2\Delta l, t)$$
$$-2(w_j^{pl,1}(L_j, t + \Delta t) - w_j^{pl,1}(L_j - \Delta l, t))$$
$$= -\frac{\lambda_j c^2}{2D_j S_j} (h(l_2^*, t_2^*) - h(l_1^*, t_1^*)) \cdot \Delta t \quad (4l)$$

Therefore, by combining (4k) and (4l), we have:

$$w_j^{pl,1}(L_j - 2\Delta l, t) - 2w_j^{pl,1}(L_j - \Delta l, t) + w_j^{pl,1}(L_j, t) =$$
$$\frac{\lambda_j c^2}{2D_j S_j} (h(l_2^*, t_2^*) - h(l_1^*, t_1^*)) \cdot \Delta t + O((\Delta t)^2). \quad (4m)$$

By expanding $h(l_2^*, t_2^*)$ in a Taylor series at $(l_1^*, t_1^*)$, we have:

$$h(l_2^*, t_2^*) - h(l_1^*, t_1^*) = o(l_2^* - l_1^*) + o(t_2^* - t_1^*) = o(\Delta l) + o(\Delta t). \quad (4n)$$

Hence, we have:

$$w_j^{pl,1}\left(L_j - \frac{2L_j}{N_j^{seg}}, t\right) + w_j^{pl,1}(L_j, t)$$
$$= 2w_j^{pl,1}\left(L_j - \frac{L_j}{N_j^{seg}}, t\right) + O((\Delta l)^2) + O((\Delta t)^2). \quad (4o)$$

Finally, we can construct the 2nd order precision numerical boundary conditions for $\pi_{j,N_j}^{pl}$ at the tail of the pipelines, as:

$$\pi_{j,N_j}^{pl} + \frac{c}{S_j} m_{j,M_j}^{pl} + \pi_{j,M_j-2}^{pl} + \frac{c}{S_j} m_{j,M_j-2}^{pl} = 2\left(\pi_{j,M_j-1}^{pl} + \frac{c}{S_j} m_{j,M_j-1}^{pl}\right). \quad (4p)$$

Similarly, we have the numerical boundary conditions for $m_{j,0}^{pl}$ at the head of the pipeline, as:

$$\pi_{j,0}^{pl} - \frac{c}{S_j} m_{j,0}^{pl} + \pi_{j,2}^{pl} - \frac{c}{S_j} m_{j,2}^{pl} = 2\left(\pi_{j,1}^{pl} - \frac{c}{S_j} m_{j,1}^{pl}\right). \quad (4q)$$

Now, we transform the PDAE system of the DEF model of IEGS into the ODAE system, in which the ODEs include (4c) and the AEs include (1c)-(1r), (4p) and (4q). In the following, we introduce the DT for the ODAEs models of IEGS.

*C. DT of the DEF Model*

*1) Transformation of NGN equations*

For the vector $Z_j(u_{j,n}^{pl})$ composed of the polynomials of variables, we use $Z_{j,n}^k$ to represent the $k$th order corresponding DT coefficient vector of the polynomials.

For the pipeline equations (4c), the initial conditions (4d), and the numerical boundaries (4p) and (4q), by applying the transformation rules 1) to 8) in Section III-A, we have:

$$(k+1)\hat{u}_{j,n}^{pl}[k+1] = V_j \frac{\hat{u}_{j,n+1}^{pl}[k] - \hat{u}_{j,n-1}^{pl}[k]}{\Delta l} + Z_{j,n}^k, \quad (5a)$$

$$j = 1, 2, \cdots, J, n = 1, 2, \cdots, N_j^{seg}$$

$$\hat{u}_{j,n}^{pl}[0] = u^{init}(x_n) \quad j = 1, 2, \cdots, J, n = 1, 2, \cdots, N_j^{seg}, \quad (5b)$$

$$\hat{\pi}_{j,M_j}^{pl}[k] + \frac{c}{S_j} \hat{m}_{j,M_j}^{pl}[k] + \hat{m}_{j,M_j-2}^{pl}[k] + \frac{c}{S_j} \hat{m}_{j,M_j-2}^{pl}[k]$$
$$= 2\left(\hat{\pi}_{j,M_j-1}^{pl}[k] + \frac{c}{S_j} \hat{m}_{j,M_j-1}^{pl}[k]\right), \quad (5c)$$

$$\hat{\pi}_{j,0}^{pl}[k] - \frac{c}{S_j} \hat{m}_{j,0}^{pl}[k] + \hat{\pi}_{j,2}^{pl}[k] - \frac{c}{S_j} \hat{m}_{j,2}^{pl}[k] = 2\left(\hat{\pi}_{j,1}^{pl}[k] - \frac{c}{S_j} \hat{m}_{j,1}^{pl}[k]\right). \quad (5d)$$

Here, we use $F_1$ to denote the pipeline equations in (5a) and $F_2 = \{F_{2,1}, F_{2,2}\}$ to denote the equations in (5c) and (5d), wherein $F_{2,1}$ represents the numerical boundary of the corresponding end of the pipelines that are connected to the nodes without the GT serving as the slack bus in EPS, and $F_{2,2}$ represents the numerical boundary of the corresponding end of the pipelines connected to the node, with the GT serving as the slack bus in EPS.

For the nodal mass flow balance equations (1c), nodal pressure consistency equations (1d) and (1e), and boundary conditions (1o) and (1n), by applying the transformation rules 3) and 4), we have:

$$\hat{m}^{nd}[k] = K^{in} \left[ \hat{m}_{1,N_1^{seg}}^{pl}[k] \cdots \hat{m}_{J,N_J^{seg}}^{pl}[k] \right]^\top \\ - K^{out} \left[ \hat{m}_{1,0}^{pl}[k] \cdots \hat{m}_{J,0}^{pl}[k] \right]^\top \quad (6a)$$

$$(K^{in})^\top \hat{\pi}^{nd}[k] = \left[ \hat{\pi}_{1,N_1^{seg}}^{pl}[k] \cdots \hat{\pi}_{J,N_J^{seg}}^{pl}[k] \right]^\top, \quad (6b)$$

$$(K^{out})^\top \text{diag}(K^{cmp})\hat{\pi}^{nd}[k] = \left[ \hat{\pi}_{1,0}^{pl}[k] \cdots \hat{\pi}_{J,0}^{pl}[k] \right]^\top, \quad (6c)$$

$$\hat{m}_n^{nd}[k] = \hat{m}^{load}[k] \quad \forall n \in S^{load}, \quad (6d)$$

$$\hat{\pi}_n^{nd}[k] = \hat{\pi}^{src}[k] \quad \forall n \in S^{src}. \quad (6e)$$

Here, we use $F_3 = \{F_{3,1}, F_{3,2}\}$ to denote the equations in (6a)-(6e), wherein $F_{3,1}$ represents the nodal equations at the nodes without the GT serving as the slack bus in EPS, $F_{3,2}$ represents the nodal equations at the node with the GT serving as the slack bus in EPS.

*2) Transformation of EPS equations*

By applying the transformation rules 3) to 5), the power flow equations (1f) to (1h) can be transformed as:

$$\hat{p}[k] = \text{diag}(\hat{e}[k]) \otimes (G\hat{e}[k] - B\hat{f}[k]) \\ + \text{diag}(\hat{f}[k]) \otimes (B\hat{e}[k] + G\hat{f}[k]), \quad (7a)$$

$$\hat{q}[k] = \text{diag}(\hat{f}[k]) \otimes (G\hat{e}[k] - B\hat{f}[k]) \\ - \text{diag}(\hat{e}[k]) \otimes (B\hat{e}[k] + G\hat{f}[k]), \quad (7b)$$

$$\text{diag}(\hat{e}[k]) \otimes \hat{e}[k] + \text{diag}(\hat{f}[k]) \otimes \hat{f}[k] \\ = \text{diag}(\hat{U}[k]) \otimes \hat{U}[k]. \quad (7c)$$

Here, we use $F_4$ to represent the equations (7a)-(7c).

By applying transformation rule 3), the boundary conditions $p$, $q$ and $U$ become $\hat{p}[k]$, $\hat{q}[k]$ and $\hat{U}[k]$, as:

$$\hat{p}_b[k] = \hat{p}^{PV}[k], \hat{U}_b[k] = \hat{U}^{PV}[k] \quad \forall b \in S^{PV}, \quad (7d)$$

$$\hat{p}_b[k] = \hat{p}^{PQ}[k], \hat{q}_b[k] = \hat{q}^{PQ}[k] \quad \forall b \in S^{PQ}, \quad (7e)$$

$$\hat{e}_b[k] = \hat{e}^{slk}[k], \hat{f}_b[k] = \hat{f}^{slk}[k] \quad \forall b \in S^{slk}. \quad (7f)$$

*3) Transformation of coupling components equations*

By applying the transformation rule 3), the coupling components equations (1i), (1j), and (1k) becomes:

$$K_{b,i}^C \hat{p}_b[k] = -K_i^{out} \left[ \hat{m}_{1,0}^{pl}[k] \cdots \hat{m}_{J,0}^{pl}[k] \right]^\top \quad \forall (i,b) \in S^C, \\ \hat{q}_b[k] = \tan\theta_b \hat{p}_b[k] \quad (8a)$$

$$\hat{p}_b[k] = -K_{b,i}^{GT} \hat{m}_i^{nd}[k] \quad \forall (i,b) \in S^{GFU}, \quad (8b)$$

$$\hat{m}_i^{nd}[k] = -K_{b,i}^{P2G} \hat{p}_b[k] \quad \forall (i,b) \in S^{P2G} \\ \hat{q}_b[k] = \tan\theta_b \hat{p}_b[k]. \quad (8c)$$

Here, we use $F_5 = \{F_{5,1}, F_{5,2}\}$ to represent the coupling components equations (8a)-(8c), wherein $F_{5,1}$ represents the equations of the GTs serving as the PV buses in EPS and the P2Gs, and $F_{5,2}$ represents the equations of the electricity-driven compressors and the GT serving as the slack bus in EPS.

***Remark 2.*** Now, the original DEF model of IEGS, (1a)-(1r), is transformed into $k$-domain system, as given in (5a)-(8c). It can be found that the $k$-domain system defines the linear recursive equations of $k$th order DT coefficients, i.e., the $k$th order DT coefficients can be obtained by solving a linear system once the 0th order to $(k-1)$th order DT coefficients are known. Therefore, through the DT-based SA algorithm, once the order $K$ is selected, we can obtain the values of the $k$th order DT coefficients from $k=1$ to $k=K$ sequentially by solving the linear equations, and then get the time-domain analytical solutions of unknowns based on (3c).

***Remark 3.*** Note that, if the semi-discrete difference scheme (4b) becomes $\partial u_j^{pl}/\partial x = (u_{j,n}^{pl} - u_{j,n-1}^{pl})/\Delta x$, the number of equations equals the number of unknown variables. However, DT of the ODE (4c) in NGN will decouple the pressure and mass flow of the gas in these equations, where the only variable in each equation is pressure or mass flow, resulting in the singularity of the coefficient matrix of the equation after DT in NGN with loop network.

## IV. SEMI-ANALYTIC SIMULATION ALGORITHM

In this section, we introduce the SA simulation algorithm tailored for the DEF model. First, we reveal the inherent block structure within the linear equations of the DT coefficients. Second, we derive a non-iterative solution method based on the block characteristics. Finally, we propose an adaptive time window control technique to adjust the DT time window by estimating the truncation error of the solution.

### A. Block Structure of DT Coefficient Equations

First, we define some new symbols to denote the DT coefficients as:

$$\hat{u}^{pl}[k] \triangleq \left[ \hat{u}_{j,0}^{pl}[k] \cdots \hat{u}_{j,N_j^{seg}}^{pl}[k] \right]_{j=1,\cdots,J}^\top \in \mathbb{R}^{\sum_{j=1}^J 2(N_j^{seg}+1)},$$

$$\hat{u}^{nd}[k] \triangleq \left[ \hat{\pi}_i^{nd}[k] \quad \hat{m}_i^{nd}[k] \right]_{i=1,\cdots,I}^\top \in \mathbb{R}^{2I}.$$

We denote $p^{cp}$ as the unknown injection active power of the buses where the GT serving as the slack bus in EPS and the electricity-driven compressors are located. Also, for variables $(\cdot)$, we define $(\cdot)[0:k] \triangleq \left[ (\cdot)[0] \quad (\cdot)[1] \cdots (\cdot)[k] \right]^\top$.

Then, we split $\hat{u}^{pl}[k]$ into three parts, denoted as $\hat{u}^{pl,1}[k]$, $\hat{u}^{pl,2}[k]$, and $\hat{u}^{pl,3}[k]$, wherein $u^{pl,1}$ denotes the DT coefficients of the states inside the pipelines, i.e., $\hat{u}^{pl,1}[k] = \left[ \hat{u}_{j,1}^{pl}[k] \quad \hat{u}_{j,2}^{pl}[k] \cdots \hat{u}_{j,N_j^{seg}-1}^{pl}[k] \right]_{j=1,2,\cdots,J}^\top$; $\hat{u}^{pl,2}[k]$ denotes the DT coefficients of the states at the corresponding end of the pipelines which are connected to the node without the GT serving as the slack bus in EPS; $u^{pl,3}$ denotes the DT coefficients of the states at the corresponding end of the pipelines connected to the node, with the GT serving as the slack bus in EPS.

The equations in IEGS (5a)-(8c) can be written in matrix form as given in (9a), wherein $b_1^k$, $b_2^k$, and $b_3^k$ are vectors whose values are all known. We calculate the DT coefficient order by order through recursion without solving the equation iteratively.

$$\begin{bmatrix} \frac{\partial F_1}{\partial \hat{u}^{pl,1}[k]} & & & & & & & \\ \frac{\partial F_{2,1}}{\partial \hat{u}^{pl,1}[k]} & \frac{\partial F_{2,1}}{\partial \hat{u}^{pl,2}[k]} & & & & & & \\ & \frac{\partial F_{3,1}}{\partial \hat{u}^{pl,2}[k]} & \frac{\partial F_{3,1}}{\partial \hat{u}^{nd,1}[k]} & & & & & \\ & & \frac{\partial F_{5,1}}{\partial \hat{u}^{nd,1}[k]} & & & & & \\ \frac{\partial F_{2,2}}{\partial \hat{u}^{pl,1}[k]} & & & \frac{\partial F_{2,2}}{\partial \hat{u}^{pl,3}[k]} & & & & \\ & & & \frac{\partial F_{3,2}}{\partial \hat{u}^{pl,3}[k]} & \frac{\partial F_{3,2}}{\partial \hat{u}^{nd,2}[k]} & & & \\ & & & & & \frac{\partial F_4}{\partial \hat{e}[k]} & \frac{\partial F_4}{\partial \hat{f}[k]} & \frac{\partial F_4}{\partial \hat{p}^{cp}[k]} \\ & \frac{\partial F_{5,2}}{\partial \hat{u}^{pl,2}[k]} & & & \frac{\partial F_{5,2}}{\partial \hat{u}^{nd,2}[k]} & & & \frac{\partial F_{5,2}}{\partial \hat{p}^{cp}[k]} \end{bmatrix} \begin{bmatrix} \hat{u}^{pl,1}[k] \\ \hat{u}^{pl,2}[k] \\ \hat{u}^{nd,1}[k] \\ \hat{u}^{pl,3}[k] \\ \hat{u}^{nd,2}[k] \\ \hat{e}[k] \\ \hat{f}[k] \\ \hat{p}^{cp}[k] \end{bmatrix} = \begin{bmatrix} b_1^k(\hat{u}^{pl}[0:k-1]) \\ b_2^k(\hat{\pi}^{src}[k], \hat{m}^{load}[k]) \\ b_3^k(\hat{e}[1:k-1], \hat{f}[1:k-1], \hat{p}[k], \hat{q}[k], \hat{U}[0:k], \hat{u}^{pl,2}[k]) \end{bmatrix}, \forall k = 1, \cdots, K \quad (9a)$$

Let us simplify (9a) in the following form:

$$\begin{bmatrix} W_{11}^k & 0 & 0 \\ W_{21} & W_{22} & 0 \\ W_{31} & W_{32} & W_{33}^k \end{bmatrix} \begin{bmatrix} \hat{x}_1[k] \\ \hat{x}_2[k] \\ \hat{x}_3[k] \end{bmatrix} = \begin{bmatrix} b_1^k \\ b_2^k \\ b_3^k \end{bmatrix}, \forall k = 1, \cdots, K, \quad (9b)$$

wherein $\hat{x}_1[k]$, $\hat{x}_2[k]$, and $\hat{x}_3[k]$ represent the DT coefficients required to solve in the corresponding three steps.

*1) Step 1: Get the internal state of the pipeline recursively*

According to the equation (5a), when calculating the internal state variables of the pipeline in step 1, $\hat{x}_1[k]$ is calculated by the recursive relation $g_1^k$, as:

$$\hat{x}_1[k] = g_1^k(\hat{u}^{pl}[0:k-1]), \forall k = 1, \cdots, K. \quad (9c)$$

wherein the relation $g_1^k$ is determined by the pipeline parameters and varies according to the order $k$. The independent variables of the recurrence relation are the lower order DT coefficients of the pipeline state. Then we write it in matrix form, as:

$$W_{11}^k \hat{x}_1[k] = b_1^k(\hat{u}^{pl}[0:k-1]), \quad (9d)$$

wherein $W_{11}^k$ is a diagonal matrix.

*2) Step 2: Obtain the state of uncoupled nodes and pipeline*

In step 2, the nodal equations $F_{2,1}$, $F_{3,1}$ and $F_{5,1}$ for nodes without the GT serving as the slack bus in EPS can be written as:

$$W_{21}\hat{x}_1[k] + W_{22}\hat{x}_2[k] = b_2^k, \quad (9e)$$

wherein $\hat{x}_1[k]$ is already solved in step 1. We consider the coefficient matrix $W_{22}$ for $\hat{x}_2[k]$ here. To simplify the calculation, only one of the nodes, $i$, is considered here. Considering that the pressures are equal at the nodes, we first use $\hat{\pi}_i^{nd,1}[k]$ to eliminate all unknown $\hat{\pi}^{pl}[k]$ in the equation, as:

$$\begin{bmatrix} V_{11} & & 1 & \\ & V_{22} & 1 & \\ -\mathbf{1}^\top & \mathbf{1}^\top & & 1 \\ & & V_{43} & V_{44} \end{bmatrix} \begin{bmatrix} \hat{m}_{out,i}^{pl,2}[k] \\ \hat{m}_{in,i}^{pl,2}[k] \\ \hat{\pi}_i^{nd,1}[k] \\ \hat{m}_i^{nd,1}[k] \end{bmatrix}, \quad (9f)$$

$$= b_{2,i}^k - W_{21,i}\hat{x}_{1,i}[k], \forall k = 1, \cdots, K$$

wherein $V_{11}$ is a diagonal matrix whose elements are $-c/S_{oi}$; $V_{22}$ is a diagonal matrix with elements $c/S_{ii}$; $oi$ and $ii$ denote the labels for all outgoing and incoming pipelines connected to node $i$, respectively; $\hat{m}_{out,i}^{pl,2}[k]$ and $\hat{m}_{in,i}^{pl,2}[k]$ are the unknown vectors of the DT coefficients about the mass flow at the head or tail of all outgoing and incoming pipelines connected to node $i$, respectively; the value of one for $V_{43}$ or $V_{44}$ indicates that the node serves as a source or load node.

If the node has the GT serving as a PV bus in EPS or P2G, and its active power is known, its coupling components equations (8b) and (8c) are in the same form as the boundary conditions of NGN, which can also be written as the fourth row in the matrix $W_{22}$. Then, we show that $W_{22}$ is of full rank. We use the first and second rows in the matrix $W_{22}$ to eliminate the first two row vectors in the left hand of the third row so that the newly formed $V_{33}$ is non-zero. Therefore, whether this node is a source node or a load node, the coefficient matrix $W_{22}$ is full rank.

In addition, we find that $W_{22}$ is only related to the parameters given in NGN, and there is no need to update during simulation.

*3) Step 3: Calculate the states of coupling nodes and EPS by non-iterative method*

Next, the equations $F_{2,2}$, $F_{3,2}$, $F_4$ and $F_{5,2}$ for EPS and the node with the GT serving as the slack bus in EPS can be written as:

$$W_{31}\hat{x}_1[k] + W_{32}\hat{x}_2[k] + W_{33}^k\hat{x}_3[k] = b_3^k \quad (9g)$$

where $\hat{x}_1[k]$ and $\hat{x}_2[k]$ is already solved in the previous steps. We consider the coefficient matrix $W_{33}^k$ for $\hat{x}_3[k]$ here.

$$\begin{bmatrix} \frac{\partial f_{2,2}}{\partial \hat{u}^{pl,3}[k]} & & & & & \\ \frac{\partial f_{3,2}}{\partial \hat{u}^{pl,3}[k]} & \frac{\partial f_{3,2}}{\partial \hat{u}^{nd,2}[k]} & & & & \\ & & \frac{\partial f_4}{\partial \hat{e}[k]} & \frac{\partial f_4}{\partial \hat{f}[k]} & \frac{\partial f_4}{\partial \hat{p}^{Cou}[k]} \\ & \frac{\partial f_{5,2}}{\partial \hat{u}^{nd,2}[k]} & & & \frac{\partial f_{5,2}}{\partial \hat{p}^{Cou}[k]} \end{bmatrix} \begin{bmatrix} \hat{u}^{pl,3}[k] \\ \hat{u}^{nd,2}[k] \\ \hat{e}[k] \\ \hat{f}[k] \\ \hat{p}^{cp}[k] \end{bmatrix} \quad (9h)$$

$$= b_3^k - W_{31}\hat{x}_1[k] - W_{32}\hat{x}_2[k], \forall k = 1, \cdots, K$$

**Algorithm 1:** DT-based Simulation of IEGS

**Input:** Total simulation time $T$, $u^{init}(x_n)$, $u^{nd}(0)$, $e(0)$, $f(0)$, $p^{cp}(0)$, boundary condition $\pi^{src}(t)$, $m^{load}(t)$, $p(t)$, $q(t)$, $U(t)$, DT order $K$;
**Output:** $\hat{u}^{pl}[0:K]$, $\hat{u}^{nd}[0:K]$, $\hat{e}[0:K]$, $\hat{f}[0:K]$, $\hat{p}^{cp}[0:K]$ at each time window;

1 **begin**
2     $t \leftarrow 0$, $\hat{u}^{pl}[0] \leftarrow u^{init}(x_n)$, $\hat{u}^{nd}[0] \leftarrow \hat{u}^{nd}(0)$, $\hat{e}[0] \leftarrow e(0)$, $\hat{f}[0] \leftarrow f(0)$, $\hat{p}^{cp}[0] \leftarrow p^{cp}(0)$;
3     Update $W_{22}$ and then calculate $W_{22}^{-1}$;
4     **while** $t < T$ **do**
5        Derive $\hat{\pi}^{src}[0:K]$, $\hat{m}^{load}[0:K]$, $\hat{p}[0:K]$, $\hat{q}[0:K]$, $\hat{U}[0:K]$;
6        Update $W_{33}$ from $\hat{e}[0]$ and $\hat{f}[0]$, and then calculate $W_{33}^{-1}$;
7        **for** $k = 1, 2, ..., K$ **do**
8            Perform Step 1: $\hat{x}_1[k] = F_1^k(\hat{u}^{pl}[0:k-1])$;
9            Update $b_2^k$ from $\hat{\pi}^{src}[k]$, $\hat{m}^{load}[k]$;
10            Update $b_3^k$ from $\hat{e}[1:k-1]$, $\hat{f}[1:k-1]$, $\hat{p}[k]$, $\hat{q}[k]$, $\hat{U}[0:k]$, $\hat{u}^{pl,2}[k]$;
11            Perform Step 2: $\hat{x}_2[k] = W_{22}^{-1}(b_2^k - W_{21}\hat{x}_1[k])$;
12            Perform Step 3: $\hat{x}_3[k] = W_{33}^{-1}(b_3^k - W_{31}\hat{x}_1[k] - W_{32}\hat{x}_2[k])$;
13        **end for**
14        Update $\Delta t$ from $\hat{u}^{pl}[K]$, $\hat{u}^{nd}[K]$, $\hat{e}[K]$, $\hat{f}[K]$, $\Delta t$;
15        $\hat{u}^{pl}[0] \leftarrow \sum_{k=0}^{K} \hat{u}^{pl}[k]\Delta t^k$, $\hat{u}^{nd}[0] \leftarrow \sum_{k=0}^{K} \hat{u}^{nd}[k]\Delta t^k$, $\hat{e}[0] \leftarrow \sum_{k=0}^{K} \hat{e}[k]\Delta t^k$, $\hat{f}[0] \leftarrow \sum_{k=0}^{K} \hat{f}[k]\Delta t^k$, $\hat{p}^{cp}[0] \leftarrow \sum_{k=0}^{K} \hat{p}^{cp}[k]\Delta t^k$;
16        $t = t + \Delta t$;
17     **end while**
18 **end**

The first two rows and the last row of $W_{33}^k$ are similar to those of $W_{22}$, except that the original boundary conditions are changed into the coupling equations. For the GT serving as the slack bus in EPS, its mass flow is calculated by the injected power of the bus. For electricity-driven compressors, the injected power of the corresponding bus is calculated from the previously calculated mass flow through the compressor by (8a). The third row of $W_{33}^k$ represents the power flow equations of EPS. As described in the review, in step 3, the EPS equations and NGN nodal equations with the GT serving as the slack bus in EPS are integrated through the coupling components equations. We increase the equations about the active power flow of the slack bus and the energy flow conversion equations of the electricity-driven compressors by the same amount as the number of unknowns for the injected active power of the slack bus and buses with electricity-driven compressors.

Furthermore, in this step, we find that $W_{33}$ is related to the 0th DT coefficient about EPS, independent of the order $k$, and only needs to be updated once in each time window.

Based on the above matrix partition, we use the non-iterative technique to solve the problem according to the characteristics of IEGS and DT. The simulation process of IEGS DEF by DT method is shown in Algorithm 1.

### B. Adaptive Time Window Control Technique

To compute the 0th order DT coefficient for the subsequent time window, we employ the adaptive window control technique [15], which is an evolution of the stepsize control strategy utilized in the Runge-Kutta method [25]. This technique dynamically adjusts the time window for each calculation based on the error tolerance we set, thereby mitigating the risk of result divergence during significant system state changes and enhancing simulation robustness.

**Algorithm 2:** DT-based Simulation of IEGS

**Input:** $K$th order DT coefficient $\hat{u}^{pl}[K]$, $\hat{u}^{nd}[K]$, $\hat{e}[K]$, $\hat{f}[K]$ as $\hat{y}[K] = [\hat{y}_1[K], \hat{y}_2[K], \ldots, \hat{y}_N[K]]$, $\Delta t$;
**Output:** $\Delta t$;

1 **begin**
2     $\tilde{y} = \hat{y}[K](\Delta t)^K$;
3     $\varepsilon_r = \text{Atol} + \min(|\hat{y}_r(0)|, |\hat{y}_r(\Delta t)|) \cdot \text{Rtol}, r = 1, 2 \ldots, N$;
4     $err = \sqrt{\frac{1}{n}\sum_{r=1}^{N}(\frac{\tilde{y}_r}{\varepsilon_r})^2}, r = 1, 2 \ldots, N$;
5     $\Delta t^{new} = \Delta t \cdot fac \cdot (1/err)^{(1/K)}$;
6     $\Delta t = \min(fac^{max} \cdot \Delta t, \max(fac^{min} \cdot \Delta t, \Delta t^{new}))$;
7 **end**

Algorithm 2 outlines the procedure for implementing the adaptive time window control technique. Initially, we estimate the truncation error $\tilde{y}$ of the $(K-1)$th order DT coefficient using the $K$th DT coefficients obtained. Subsequently, we determine the error tolerance $\varepsilon$ by considering the absolute tolerance Atol, relative tolerance Rtol, and the variable values within the time window. The root mean square error $err$ is then calculated using $\tilde{y}$ and $\varepsilon$, and the size of the time window is adjusted under the constraint of the change rate. Wherein $fac$ represents a conservative factor less than one; $fac^{max}$ and $fac^{min}$ serve as upper and lower bounds to prevent abrupt changes in temporal steps; $\hat{y}[K]$ denotes the vector containing all $N$ DT coefficients of the $K$th order.

**Remark 4.** When calculating the DT coefficients of each order for the states between nodes in NGN during step 2, we observe that these coefficients are decoupled by the pipeline and do not mutually influence each other. This decoupling arises because the DT coefficients of the internal state of the pipeline have already been computed in step 1, thus preventing the nodes from interacting through the pipeline. However, it is important to note that this decoupling is specific to the calculation of a particular order of the DT coefficient. The changes in the node state variables remain coupled within the time window. This coupling occurs because the computation of subsequent order DT coefficients involves the DT coefficient of the internal state of the pipeline, which is influenced by the DT coefficient at the boundary of pipelines, which in turn is related to the node state.

### V. CASE STUDIES

In this section, the proposed non-iterative method is tested on two systems: 1) a 50 km single gas pipeline; 2) a 133-node-118-bus system. For comparative analysis, the following methods are employed: 1) M1: Characteristic line method solved using Newton's method, serving as the benchmark for accuracy analysis; 2) M2: Implicit Euler difference scheme solved using Newton's method; M3: Implicit central difference scheme solved using Newton's method; M4: DT method with 5th order.

In the first system, we demonstrate the superior accuracy of the proposed method compared to the numerical methods. In the second system, we validate the increased computational efficiency of the proposed method over the numerical methods. All tests are conducted using Python 3.11.9 on a computer with an Intel i9-13900K CPU and 96 GB of RAM.

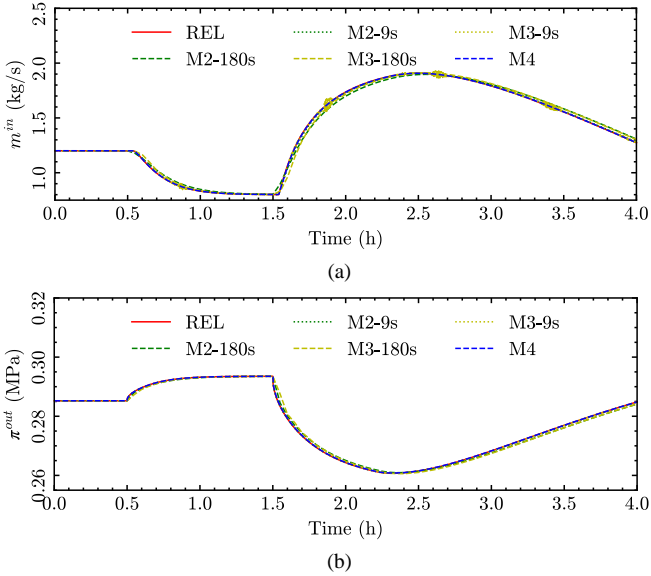

Fig. 2. Accuracy comparison on a single pipeline system: (a) $m^{in}$; (b) $\pi^{out}$.

TABLE I
RMSEs AGAINST REFERENCE DATA

| Variable | M2-180s | M2-9s | M3-180s | M3-9s | M4 |
|---|---|---|---|---|---|
| $m_0$ | 2.70e-3 | 1.48e-4 | 2.78e-3 | 2.18e-4 | 1.66e-5 |
| $p_N$ | 6.26e1 | 1.01 | 8.83e1 | 6.90e-1 | 1.76 |

### A. A 50km Single Gas Pipeline System

The single gas pipeline system comprises a pipeline linking the source and load nodes [3]. The source node pressure is fixed at 300 kPa, whereas the load mass flow varies: it decreases from 1.2 kg/s to 0.8 kg/s when time reaches 0.5 hours, then increases to 2 kg/s at 1.5 hours, and decreases gradually after 2 hours.

The results obtained from M1 are utilized as reference data to assess the accuracy of M2-M4. A uniform spatial step of 1 km is adopted for all simulations. For M2 and M3, the tests are performed at temporal steps of 180 seconds and 9 seconds. The adaptive time window control technique is employed in M4 to ensure that the temporal step yields the results within the acceptable error tolerance, thereby enhancing the robustness of the proposed method against significant disturbances.

As we can see from Fig. 2, the results obtained by M2 and M3 with a small temporal step and M4 are close to the reference data generated by M1, indicating the high accuracy achieved by these methods. In contrast, M2 and M3 with a larger temporal step exhibit a more significant error. Table I presents each method's root mean square errors (RMSEs) compared to the reference data to quantify the accuracy differences. This quantitative assessment yields a notable observation: Numerical methods require considerably smaller temporal steps to attain accuracy levels comparable to M4. Although refining the temporal discretization enhances accuracy, it substantially increases computational costs and execution time, rendering it challenging for real-time applications or large-scale simulations. Consequently, M4 stands out as an effective and practical choice for addressing complex integrated energy network problems, owing to its inherent ability to maintain high accuracy.

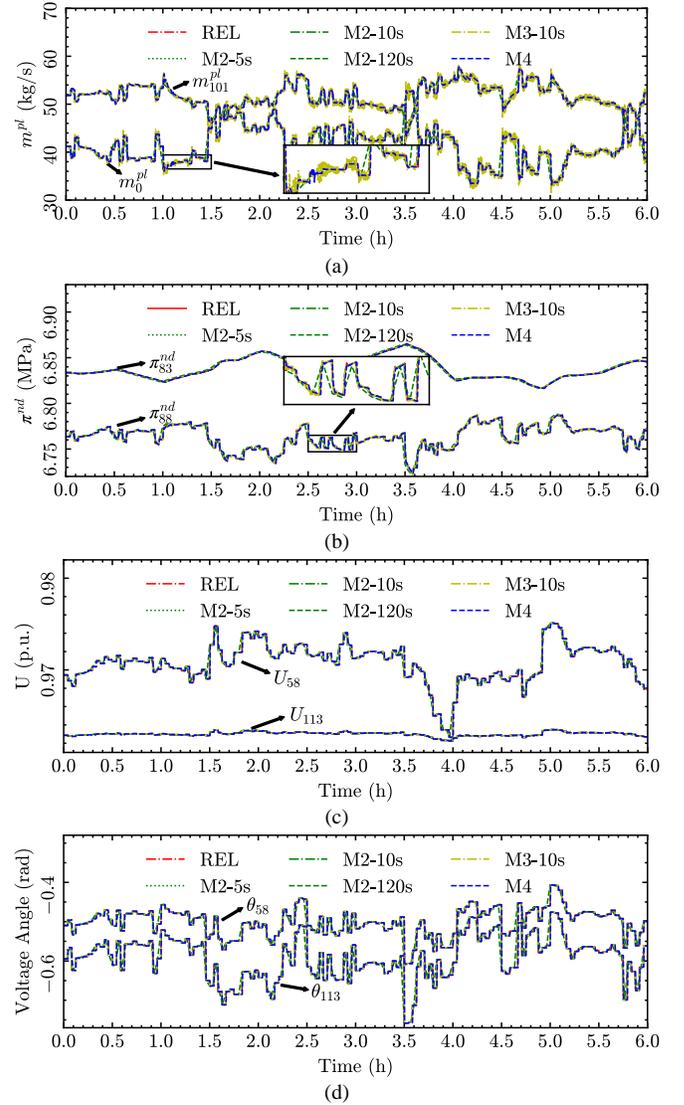

Fig. 3. Accuracy comparison on a 133-node-118-bus system: (a) $m_0^{pl}$ and $m_{101}^{pl}$; (b) $\pi_{88}^{nd}$ and $\pi_{83}^{nd}$; (c) $U_{58}$ and $U_{113}$; (d) $\theta_{58}$ and $\theta_{113}$.

### B. A 133-node & 118-bus System

The effectiveness and efficiency of the proposed method are validated in a large-scale system, which is an improved system of the 133-node NGN and IEEE 118-bus EPS from [26, 27]. The coupling components in the system consist of 5 GTs, a P2G, and an electricity-driven compressor. Specifically, the GT serving as the slack bus at bus 69 in EPS is supported by NGN at node 89. The remaining 4 GTs, serving as PV buses at buses 10, 12, 49, and 61 in EPS, are supported by NGN at nodes 133, 130, 90, and 98, respectively. The P2G at node 79 in NGN is supplied by EPS at bus 59, and the electricity-driven compressor at node 84 in NGN is powered by EPS at bus 114.

Utilizing the results obtained from M1 with a spatial step of 500 m as a baseline, the calculation accuracy and speed of methods M2-M4 are evaluated under the same spatial step of 1000 m.

As shown in Fig. 3, M2-M4 effectively capture the dynamic changes in mass flow and pressure in NGN, and voltage distribution in EPS. Particularly for the mass flow and pressure states in NGN, the results of M2 and M3 with smaller temporal steps and M4 exhibit good agreement with the reference results. However, M2 and M3 with a large temporal step deviate substantially when the system boundary changes significantly. We can find from Table II that the calculation accuracy of M4 in EPS is comparable to that of M2 and M3. Notably, the accuracy of M4 in NGN surpasses that of M2 and M3 with similar temporal steps. Furthermore, despite M3 with 2nd order precision in the difference scheme, numerical oscillations occur during the simulation process, leading to lower mass flow calculation accuracy than M2 with 1st order precision.

TABLE II
RMSEs OF DIFFERENT METHODS

| Variable | M2-120s | M2-10s | M2-5s | M3-10s | M4 |
|---|---|---|---|---|---|
| $e$ | 1.614e-2 | 8.222e-3 | 3.357e-3 | 8.222e-3 | 4.242e-3 |
| $f$ | 2.741e-2 | 1.390e-2 | 5.674e-3 | 1.390e-2 | 7.199e-3 |
| $m^{pl}$ | 3.605e-1 | 8.950e-2 | 6.509e-2 | 1.299e-1 | 9.317e-2 |
| $p^{nd}$ | 1.378e3 | 3.047e2 | 1.157e2 | 3.158e2 | 1.743e2 |

TABLE III
COMPUTATIONAL COST OF DIFFERENT METHODS

| Methods | M2-120s | M2-10s | M2-5s | M3-10s | M4 |
|---|---|---|---|---|---|
| Time Cost(s) | 6.7 | 63.7 | 126.6 | 84.6 | 56.8 |
| Step Number | 180 | 2160 | 4320 | 2160 | 2586 |

As presented in Table III, a comparative analysis of the computational cost among multiple methods reveals further advantages of M4 over numerical methods such as M2 and M3. The advantage of M4 lies in its utilization of recursive calculations and linear equation solving, eliminating the requirement for iterative solutions. As a result, when using comparable temporal steps, M2 and M3 demonstrate longer total calculation times than the proposed M4. Previous findings show that M2 and M3 require considerably smaller temporal steps to achieve an accuracy level comparable to M4. Furthermore, it is well-established that, for the given method, calculation time is directly proportional to the number of calculation steps and inversely proportional to the size of the temporal step. This relationship further confirms the superior computational efficiency of the proposed M4 method.

## VI. CONCLUSION

In this paper, we have addressed the challenges associated with DEF simulation in IEGS by proposing a novel non-iterative SA method based on DT. Our method achieves high accuracy with a small computational cost by leveraging the recursive properties of DT. Simulation results demonstrate the effectiveness and superiority of the proposed approach.